\documentclass[twocolumn]{article}
\pdfoutput=1
\usepackage{layouts}
\usepackage[margin=1in]{geometry}
\usepackage{blindtext}
\usepackage{soul}
\usepackage{abstract}
\usepackage{graphicx}
\usepackage{graphbox}
\usepackage{authblk}
\usepackage{hyperref}

\usepackage{cuted}
\usepackage{subcaption}
\usepackage{caption}
\usepackage{siunitx}
\usepackage{amsmath}
\usepackage{amssymb}
\usepackage{enumitem}
\usepackage{placeins}
\newcommand{\low}{\textsc{low}}
\newcommand{\high}{\textsc{high}}
\usepackage[noadjust]{cite}
\title{Logic Gates based on Interaction of Counterpropagating Light in~Microresonators }
\author[1 2 3]{Niall~Moroney}
\author[1 2 4]{Leonardo~Del Bino}
\author[2 4]{Michael~T.~M.~Woodley}
\author[1 2 3]{George~N.~Ghalanos}
\author[2 5]{Jonathan~M.~Silver}
\author[2 3]{Andreas~\O ~Svela}
\author[1 2]{Shuangyou~Zhang}
\author[1 2 *]{Pascal~Del'Haye}
\affil[1]{Max Planck Institute for the Science of Light (MPL), Erlangen 91058, Germany}
\affil[2]{National Physical Laboratory (NPL), Teddington TW11 0LW, United Kingdom}
\affil[3]{Imperial College London, SW7 2AZ, United Kingdom}
\affil[4]{Heriot-Watt University, Edinburgh, EH14 4AS, United Kingdom}
\affil[5]{City, University of London, London, EC1V 0HB, United Kingdom}
\date{}
\affil[*]{Corresponding Authors: Pascal Del'Haye, pascal.delhaye@mpl.mpg.de}

\begin{document}

\maketitle

\begin{abstract}
Optical logic has the potential to replace electronics with photonic circuits in applications for which optic-to-electronic conversion is impractical and for integrated all-optical circuits. Nonlinear optics in whispering gallery mode resonators provides low power, scalable methods to achieve optical logic. We demonstrate, for the first time, an all-optical, universal logic gate using counterpropagating light in which all signals have the same operating optical frequency. Such a device would make possible the routing of optical signals without the need for conversion into the electronic domain, thus reducing latency. The operating principle of the device is based on the Kerr interaction between counter-propagating beams in a whispering gallery mode resonator which induces a splitting between the resonance frequencies for the two propagating directions. Our gate uses a fused silica microrod resonator with a \textit{Q}-factor of $\SI{2e8}{}$. This method of optical logic gives a practical solution to the on-chip routing of light.
\end{abstract}

\section{Introduction}
Fibre optic technology has underpinned the development of internet communications over the last few decades allowing the transmission of vast amounts of information with reduced latency. However, there are fears that internet latency and bandwidth limitations will slow progress in developing real-time applications, particularly for what is often referred to as the internet of things. Accordingly, methods to increase internet speeds are of great interest and one proposed solution is to reduce the latency associated with optic-to-electronic conversion at network nodes by keeping the signal in the optical domain \cite{OpticalNetworks2,OpticalNetworks3}. A photonic processor will instead be used to route the incoming signal to the correct output port \cite{OpticalNetworks}, utilising optical logic gates in the process. Whispering gallery mode (WGM) resonators \cite{GuopingLin2017} are a promising candidate for optical logic gate architectures as their high \textit{Q}-factors and small mode volumes \cite{Q-Factor} allow for the required nonlinear optical phenomena for only modest input powers \cite{Applications,subMW}. Furthermore, such devices can be integrated on-chip using CMOS technology and made from a wide range of materials, ensuring scalability \cite{CMOS1,CMOS2,CMOS3}. Optical logic gates have been previously demonstrated in WGM resonators, but with associated  issues that would prevent easy integration into optical networks. These issues include: requiring multiple operation frequencies \cite{Xu2007,WGM_Recent,Sethi2014,Similar,FWM,FWM2,TwoPhoton,Wavelength1,Wavelength2}, needing electronic control \cite{Zhang2010, Reconfigurable, Electronic, Heater,Thermal} or requiring pulsed inputs \cite{Ibrahim2004}.
\par
Here we present an all-optical, universal logic gate in which all signals operate at the same frequency in the telecom band. Such a device could be an important step in the development of scalable cascaded logic for the all-optical routing of optical signals.

\section{Concept}
The optical Kerr effect leads to an intensity dependent refractive index that becomes appreciable for suitably intense input light \cite{NonlinearOptics,OpticalKerr}. In particular, for counter-propagating light, the refractive index change is mediated by self-phase modulation (SPM) -- where the beam's own intensity causes the change -- and cross phase modulation (XPM) -- where the intensity of the counter-propagating light causes the change. Note that in this work XPM is induced by a counterpropagating light wave of the same frequency in contrast to the more common case of XPM induced by a co-propagating light wave with a different frequency. Importantly, for dielectrics, XPM has twice the effect of SPM, meaning two counter-propagating beams of different intensities will experience different effective refractive indices \cite{Kaplan1981}:
\begin{equation}
\Delta n_{1,2} \propto n_2 \left(I_{1,2} + 2 I_{2,1} \right)
\label{DeltaN}
\end{equation}
where the subscripts $i \in \{ 1,2 \}$ denotes the two propagation directions clockwise (CW) and counterclockwise (CCW) respectively, $\Delta n_i$ are the intensity-dependent refractive index changes, $n_2$ is the nonlinear refractive index, and $I_i$ are the propagating field intensities.
\par
For two counter-propagating inputs of dimensionless powers $\tilde{p}_i$, at cold-cavity detunings of $\Delta_i$, assuming lossless coupling from the input waveguide, the circulating field powers, $p_i$, can be shown to obey (See Appendix \ref{Normalisation} for the normalisation process) \cite{DelBino2017}:
\begin{equation}
p_{1,2} = \frac{\tilde{p}_{1,2}}{1 + \left(p_{1,2} + 2p_{2,1} - \Delta_{1,2} \right)^2} 
\label{CircPower}
\end{equation}
This phenomenon leads to a difference between the resonance frequencies for both propagating directions. When pumped equally ($\tilde{p}_1 = \tilde{p}_2$, $\Delta_1=\Delta_2$), this resonance splitting leads to spontaneous symmetry breaking for appropriate input powers and detunings \cite{DelBino2017, Francois}. This effect has been used to demonstrate optical isolators and circulators \cite{DelBino2018}.
\par
\begin{figure}[t]
	\centering
%
%
\includegraphics[width=\linewidth]{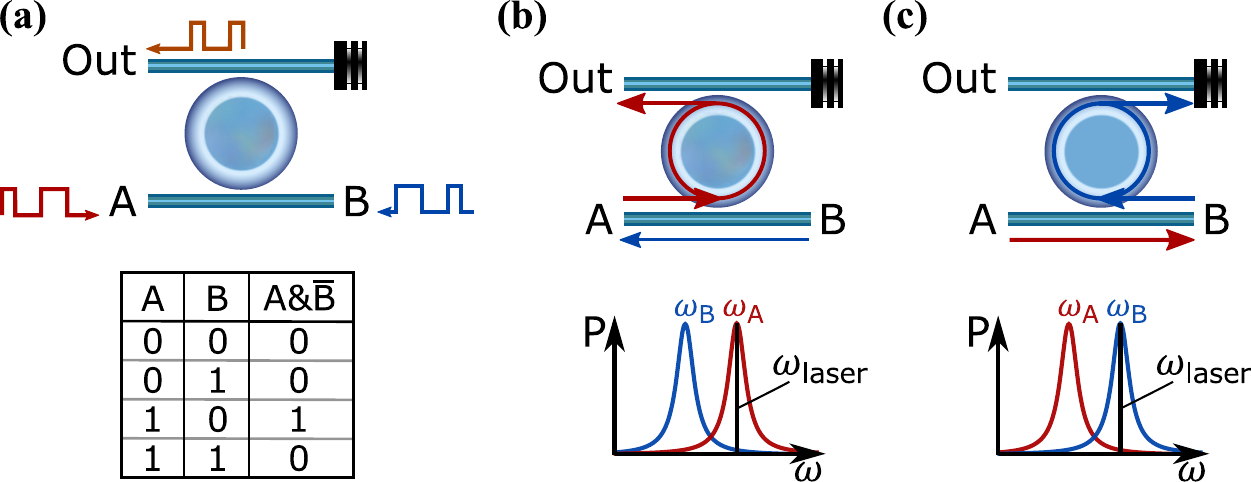}
  \caption{Logic gate concept. \textbf{(a)} Two amplitude modulated input beams, $A$ and $B$, are coupled into a WGM resonator via a waveguide. A second waveguide couples the light to the output port, with a beam blocker on the remaining unused port. Nonlinear interactions of the counterpropagating beams, mediated by the Kerr effect, ensure that the output power is only \high\ when input $A$ is \high\ and input $B$ is \low. The associated truth table is shown, which is equivalent to the universal gate $A\&\bar{B}$. \textbf{(b)} For sufficiently high input powers, the nonlinear refractive index leads to a splitting between the resonance frequencies in the clockwise and counter-clockwise directions. The input laser can be tuned such that when input power at $A$ is higher than $B$, the resonator only supports CCW propagation and consequently a \high\ output, satisfying the third row of the truth table. \textbf{(c)} Likewise, when the input power at $B$ is higher than $A$, the CCW mode's resonance frequency is shifted such that $A$ does not couple through the resonator and the output power is \low , satisfying the fourth row of the truth table. Trivially the top two rows of the truth table are satisfied -- when $A$ is \low , the output is \low.}
  \label{fig:basicIdea}
\end{figure}

Here we deliberately pump the system asymmetrically to achieve the operation of a universal gate -- the $A \& \overline{B}$ -- which can be used to realise any boolean computation when suitably cascaded, allowing for the simple design of complex logical devices. Fig.~\ref{fig:basicIdea}(a) shows the operating principle and truth table for a $A \& \overline{B}$ gate: the inputs $A$ and $B$ correspond to beams coupled into the resonator, in opposite directions, via an input waveguide. A second waveguide is then used to out-couple the output of the gate which is only \high\ when input $A$ is \high\ and input $B$ is \low; the output is otherwise \low .
\par
Trivially, the top two rows of the truth table are satisfied because the output is taken from the resonator's CCW propagating field which in turn is fed from input $A$. Accordingly when $A$ is \low\, the output is \low. 
\par
When the input fields are resonant with the cavity, the intracavity field will build up and this resonance condition is dependent on the refractive index of the medium due to its effect on the field's wavelength. However, (\ref{DeltaN}) shows that two counter-propagating fields of different intensities experience different refractive indices. Accordingly, there is a difference in the resonance frequencies for both propagation directions due to the Kerr nonlinearity, as shown in Fig.~\ref{fig:basicIdea}(b,c).  
\par
Figure \ref{fig:basicIdea}(b) shows the case when the intensity of input $A$ is greater than that of $B$.  There is a difference in the resonance frequencies of the two propagation directions, with the CW (blue) resonance being red-shifted further than the CCW (red) due to the larger refractive index shift from XPM. This has caused the input $A$ to become resonant with the cavity, with a large proportion of its power being transmitted through the cavity to the output, whilst the input $B$ is far detuned from its highly shifted resonance leading to a significantly reduced coupling efficiency. For Fig.~\ref{fig:basicIdea}(c), the intensity of input $B$ is greater than $A$ and the opposite outcome occurs. $A$ is now far detuned from its shifted resonance and so the output is well suppressed. Together, these allow the realisation of the bottom two rows of the truth table provided input $B$ has some positive power offset in comparison with $A$. When $A$ is \high\ and $B$ is \low\ (row 3), we have the same situation as shown in Fig.~\ref{fig:basicIdea}(b) and the output is \high. However, when both $A$ and $B$ are \high\ (row 4), the positive offset ensures that the output is suppressed, as per Fig.~\ref{fig:basicIdea}(c). 
Accordingly, all rows of the truth table are satisfied and thus this system operates as an $A \& \overline{B}$ gate. Also worthy of note is that this concept works regardless of the choice of input ports, so long as both inputs counter-propagate. Accordingly, this gate can be reconfigured to work in different directions with minimal changes required. 

\section{Experimental Setup}
We use a $1.5$~$\si{\milli \meter}$ diameter, high-$Q$ WGM microrod resonator that has been machined from fused silica in a CO\textsubscript{2} laser lathe \cite{LaserLathe}. Subsequent surface reflow of the resonator, performed by reducing the CO\textsubscript{2} laser power, improves the surface smoothness, giving a quality factor $Q_0 \simeq \SI{2e8}{}$.
\par
\begin{figure}[h]
  \includegraphics[width=\linewidth]{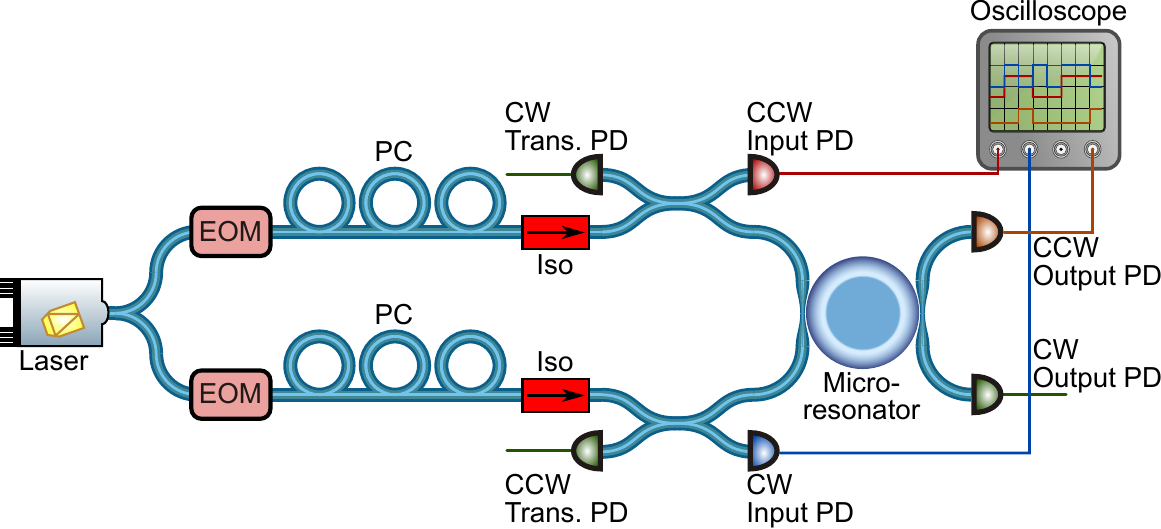}
  \caption{Experimental setup. A laser beam is split into two branches, corresponding to inputs $A$ and $B$. Each input is independently amplitude modulated by a Mach-Zender electro-optic-modulator (EOM) and has its polarisation matched to that of the resonant mode using a polarisation controller (PC). They are then coupled in opposite directions into a fused silica microrod resonator via a tapered fibre, with directional couplers and photodiodes used to monitor the power input into and transmitted through the tapered fibre in both directions. A second tapered fibre is used to couple out the resonator field, and is attached to photodiodes to monitor the output.}
  \label{fig:setup}
\end{figure}

Light from an amplified $1.55$~$\si{\micro \meter}$ laser source is split and coupled into both directions of the resonator via a tapered optical fibre, see Fig.~\ref{fig:setup}. A second tapered fibre is used to out-couple the cavity field and is attached to photodiodes to monitor the logic gate output \cite{CLEO}.
\par
Both input branches are amplitude modulated by a fibre-coupled Mach-Zender electro-optic modulator (EOM) before having their polarisation matched to that of the resonator mode to maximise the in-coupled power. Directional couplers and photodiodes are used to monitor the powers input and transmitted through the tapered fibre in each direction, and optical isolators ensure that the signals do not return to the laser.
\par
Custom waveforms are sent from a function generator to the EOMs, switching inputs $A$ and $B$ between \high\ and \low\ values, representing the logical inputs $1$ and $0$ respectively. Input $A$ is modulated between $0$ and $35$~$\si{\milli \watt}$, whereas input $B$ is modulated between $15$ and $50$~$\si{\milli\watt}$. The bias on input $B$ of $15$~$\si{\milli\watt}$ is required to force the output into the \low\ state when both inputs are \high. 

\section{Results}
\begin{figure}[h]
\includegraphics[width=\linewidth]{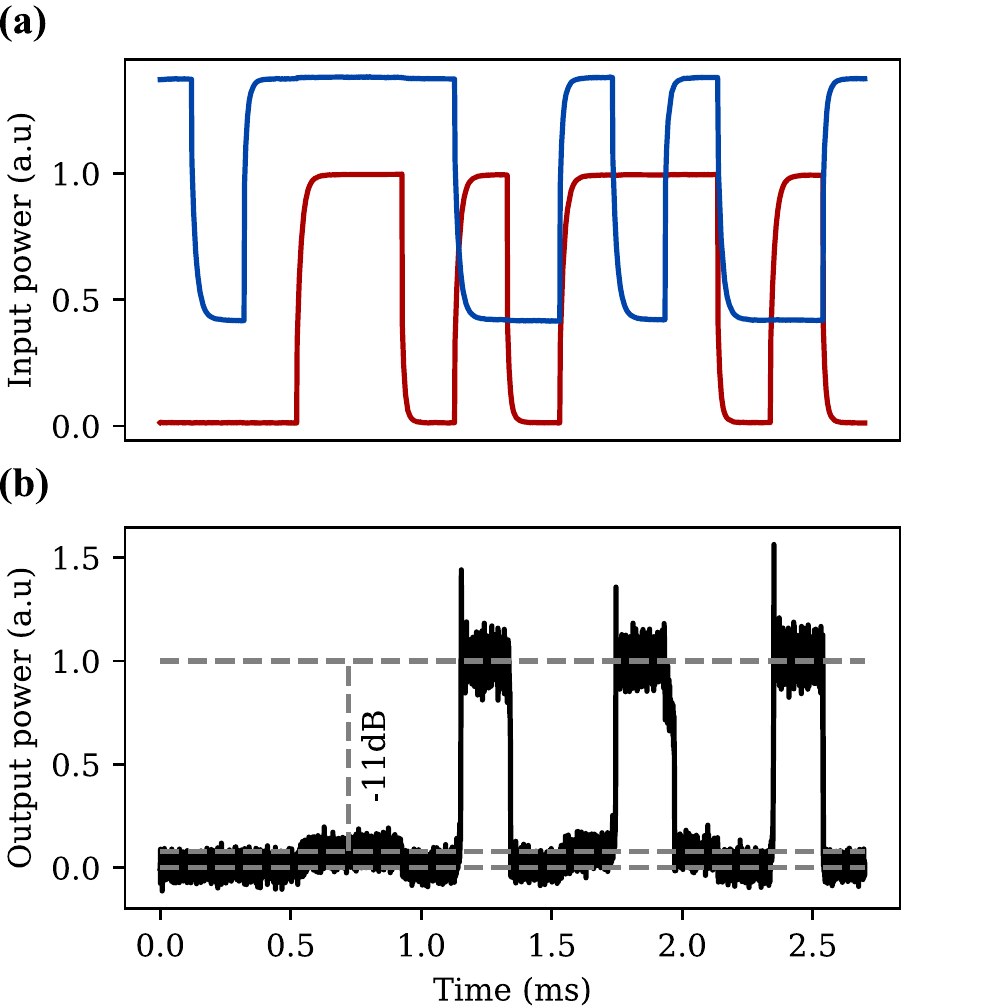}
  \caption{Demonstration of the all-optical logic gate. Panel \textbf{(a)} shows the measured powers input into the tapered fibres in both directions. Inputs $A$ and $B$ are both amplitude modulated by the same amount, but with $B$ having a positive offset in order to suppress the output when both inputs are \high. Panel \textbf{(b)} shows the measured output field, which can be seen to be \high\ only when input $A$ is \high\ and input $B$ is \low, showing the correct operation of an  $A \& \overline{B}$ gate. When both inputs are \high\ the output is not fully suppressed, which can be seen by the slight increase between $0.5$ and $1$~$\si{\milli \second}$. This residual power and the proper output power together characterise the suitability of the logic gate to be cascaded, with a low residual power and a high signal power being optimal.}
  \label{fig:results}
\end{figure}

Figure \ref{fig:results} shows the photodiode outputs for an experimental run. Fig.~\ref{fig:results}(a) shows that inputs $A$ and $B$ were modulated by the same amplitude, but with $B$ having a positive offset in comparison to $A$, to ensure output suppression when both inputs are \high.
\par
Figure~\ref{fig:results}(b) shows the output of the logic gate. It correctly gives a \high\ output only when the input $A$ is \high\ and input $B$ is \low. The laser frequency was tuned in from the blue side of the resonance until this correct signal was observed, at which point the resonator was thermally locked to the laser and no active control was required to maintain correct operation.
\par
There is a qualitative difference between the output state change at $\sim 1.8$~$\si{\milli \second}$ and the changes at $\sim 1.4$ and $2.6$~$\si{\milli \second}$. The sharp state changes are due to the associated input being turned off, whereas the state change at $\sim 1.8$~$\si{\milli \second}$ is mediated solely by the nonlinear interaction between the counter-propagating beams, causing this qualitative difference. 
\par
The output is suppressed by $11$~$\si{\decibel}$ when both inputs are \high. This suppression parameter is an important measure of how suitable the system is for cascaded operation as subsequent gates need to distinguish between a ``true" and a ``false" \high\ power and is within a range that has previously been predicted for optical logic gates \cite{FWM2}. 
\par
For the measurement, the output waveguide was significantly undercoupled to the cavity in order to maximise the nonlinear suppression effects for this proof of principle, meaning only a small fraction of the intracavity field was output to the photodiode. For these low powers, electronic noise in the photodoide is visible and does not indicate intensity noise in the output beam.

\section{Performance Analysis}
The performance of the logic gate can be characterised by two parameters: total output attenuation and \low\ state suppression. This section will describe the theoretical limits of the device performance.
\par
A dimensionless time-dependent model is used to calculate the circulating electric fields, $e_i$ \cite{Woodley2018}:
\begin{equation}
\dot{e}_{1,2} = \tilde{e}_{1,2} - \left[ 1 + i \left(  |e_{1,2}|^2 + 2|e_{2,1}|^2 - \Delta_{1,2} \right)  \right]e_{1,2}
\label{TimeDependent}
\end{equation}
where $\tilde{e}_i$ and $\Delta_i$ are the input electric field and input detuning from the cavity resonance in the $i$ direction respectively ($i=1\:(2)$ corresponds to the CCW (CW) direction). Note that, as the input fields are from the same source, $\Delta_1 = \Delta_2$. The circulating and input electric fields are related to their respective dimensionless powers by $p_i = |e_i|^2$ and $\tilde{p}_i = |\tilde{e}_i|^2$.
\par
The input powers are given in terms of the modulation amplitude and offset as:
\begin{subequations}
\begin{align}
\tilde{p}_1 &= P_m A \\
\tilde{p}_2 &= P_m \left(B + \zeta \right)
\end{align}
\end{subequations}
where $A, B \in \left \{0, 1 \right \}$ are the logical inputs, $P_m$ is the dimensionless input modulation amplitude and $\zeta \in \left ( 0,1 \right)$ is the modulation offset as a proportion of $P_m$. 
\par
An ideal logic gate will have an output power of $P_m$ when $A=1, B=0$. This requires a resonance condition for this situation, which, examining (\ref{CircPower}), gives the following detuning condition:
\begin{equation}
\Delta_{1,2} = P_m + 2 p_{2,\text{low}}
\label{Detuning Condition}
\end{equation}
where $p_{2,\text{low}}$ is the corresponding circulating CW field which can be found as the solution to the cubic equation:
\begin{equation}
p_{2,\text{low}} \left(1+ \left(P_m - p_{2,\text{low}} \right )^2 \right) - \zeta= 0
\end{equation}
With this condition satisfied, the transmission efficiency is limited by the intrinsic round-trip losses in the resonator, which are small for the high-$Q$ resonators here investigated, and by the power lost via the through-port. The latter can be minimised to zero in the limit of high over-coupling for both waveguides, though this would degrade the effective $Q$ of the resonator \cite{AddDropFilter}.
\par
The \low\ state suppression was then analysed by numerically evaluating the steady state solution of (\ref{TimeDependent}). Inputs $A$ and $B$ were switched between $0$ and $1$ such that every state change was observed, for $P_m$ and $\zeta$ values. The output signal was then recorded, normalised by $P_m$, to find the suppression of the \low\ state. For all simulations, the detuning was set to maximise the \high\ state as per (\ref{Detuning Condition}).
\par
\begin{figure*}[h]
  	\includegraphics[width=\linewidth]{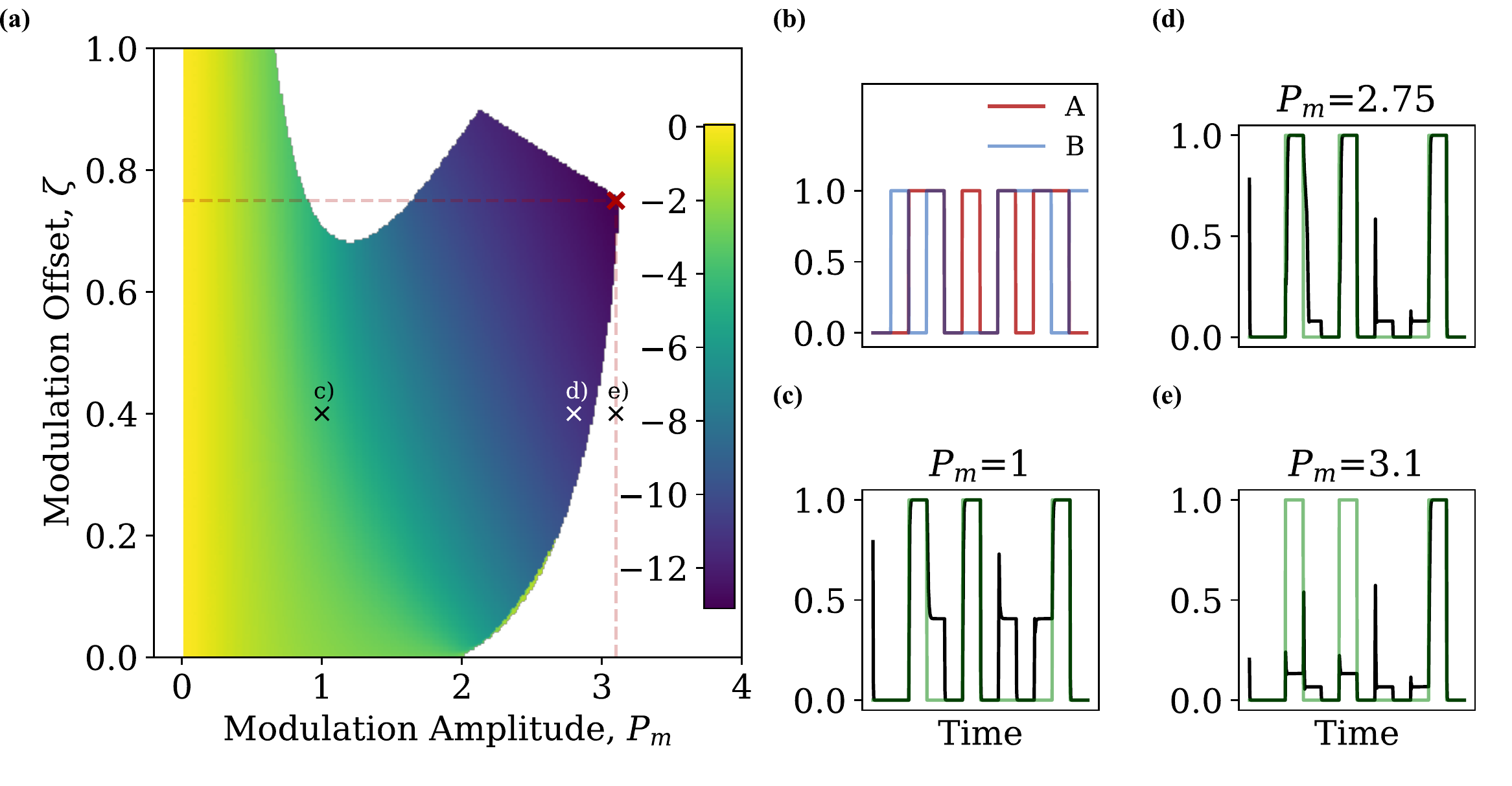}
  \caption{Performance Analysis. \textbf{(a)} Suppression of \low\ output state for different dimensionless modulation amplitudes and offsets in $\si{\decibel}$. Low values indicate suitability for cascaded operation, which is improved for higher input modulation amplitudes and offsets up to a maximum of $13$~$\si{\decibel}$ at $P_m \approx 3.1$, $\zeta \approx 0.75$ (red cross). The whitespace shows the regions for which the inability to correctly access all states inhibits the operation of the logic gate. \textbf{(b)} Inputs $A$ and $B$ for the timestep simulation presented in (c)-(e). \textbf{(c)-(e)} Simulated output for different modulation amplitudes at a modulation offset of $\zeta=0.4$. For low amplitude (c) the output is only suppressed to roughly $3$~$\si{\decibel}$ of its maximum value. This suppression improves with a higher modulation amplitude (d), but eventually the system stops behaving suitably (e): for sufficiently large input powers there are instances where the output state should be \high\, but it cannot reach this state because of hysteresis in the system.}
  
\label{fig:Theory}
  
\end{figure*}

\noindent Figure \ref{fig:Theory}(a) shows the results of this simulation: a maximum \low\ state suppression of $13$~$\si{\decibel}$ is attained when $P_m \approx 3.1$, $\zeta \approx 0.75$, meaning our demonstration of a suppression of $11$~$\si{\decibel}$ is near optimal. Figs.~\ref{fig:Theory}(c)-(e) show the results of the output from the time-dependent model for inputs shown in Fig.~\ref{fig:Theory}(b) and modulation parameters shown in Fig.~\ref{fig:Theory}(a). Fig.~ \ref{fig:Theory}(c) exhibits only a modest suppression of the signal when $A=1$ and $B=1$, however this is improved for the higher $P_m$ in Fig.~\ref{fig:Theory}(d).
\par
Excess power in the resonator leads to a hysteresis effect that inhibits the correct logical operation. This is shown in Fig.~\ref{fig:Theory}(e) where two \high\ states have not been accessed, with similar failures present for all of the whitespace of Fig.~\ref{fig:Theory}(a). It is due to this hysteresis induced limitation that the input bit sequence shown in Fig.~\ref{fig:results} was used. This bit sequence includes all possible input state switches and the correct output signal shows that this logic gate will work for arbitrary input bit sequences. This ensures that the logic gate can work with no dead-time in the input signals, ensuring suitability for continuous data processing applications (though arbitrarily high \low\ state suppressions are available if the device were to operate on a return to zero input scheme as there is no associated hysteresis induced power limit) \cite{Similar}.  
\par
The final performance characteristic of this device is the speed at which it can operate. Here we have demonstrated operation at $5$~kbps, but the observed response times of $< 60$~$\si{\micro \second}$ indicates that a bitrate of $16$~kbps could be achieved with our setup. These switching speeds are limited by the cavity lifetime, which is large for the high-\textit{Q} cavity used, leading to a compromise regarding the \textit{Q}-factor: high \textit{Q}-factors allow low-power operation but the associated long cavity lifetime limits the operation speed \cite{Pollinger2010,Similar}. Using a smaller resonator with a higher nonlinearity can mitigate this effect as a lower $Q$-factor is needed for the same input power. A proposed silicon-nitride-resonator-based logic gate would operate at sub-$\si{\milli \watt}$ input powers with a bitrate of the order of Gbps. An extensive study into the switching dynamics of this system is given in \cite{SwitchingPaper}.
\section{Conclusion}
We have demonstrated all-optical, universal logic gates based on the Kerr interaction of counter-propagating light in a WGM resonator. This is the first demonstration of such a device in which all signals operate at the same optical frequency, making it a promising candidate for cascaded operation in the all-optical routing and information processing of optical signals. We demonstrate a device that operates, for arbitrary input signals at a power difference of $11$~$\si{\decibel}$ between output \high\ and \low\ states. Integration of such a device on-chip with a suitably highly nonlinear resonator would allow for  Gbps operation for sub-$\si{\milli \watt}$ input powers, with arbitrarily high throughput efficiencies available with increasing input powers. In particular, with improved microfabrication techniques, the presented optical logic gates based on Kerr-interaction in microresonators could pave the way for novel types of photonic integrated circuits.

{\small \textit{This work was supported by the following grants: H2020 Marie Sklodowska-Curie Actions (MSCA) (748519, CoLiDR); Horizon 2020 Marie Sklodowska-Curie grant (GA-2015-713694); National Physical Laboratory Strategic Research; H2020 European Research Council (ERC) Starting Grant (756966, CounterLight); N.~M. and A.~\O.~S. acknowledge funding from the Engineering and Physical Sciences Research Council (EPSRC) via the Quantum Systems Engineering Skills Hub; L.~D.~B. and M.~T.~M.~W. acknowledge funding from the EPSRC via the CDT for Applied Photonics. G.~N.~G. acknowledges funding from the EPSRC via the CDT for Controlled Quantum Dynamics; J.~M.~S. acknowledges funding from the Royal Academy of Engineering and the Office of the Chief Science Advisor for National Security under the UK Intelligence Community Postdoctoral Fellowship Programme.}}

\appendix
\section{Normalisation process}
\label{Normalisation}

The dimensionless parameters herein used to describe the changing field intensities are related to the corresponding dimensional parameters of a critically coupled system as follows \cite{Woodley2018}:
\begin{enumerate}[label=(\alph*)]

\item $\omega_0$ is the cold cavity resonance frequency
\item $\omega_{\textrm{las},i}$ is the input laser frequency in direction $i$
\item $\gamma_0$ is the intrinsic half linewidth of the resonance
\item $\kappa$ is the linewidth associated with coupling to the cavity
\item $n_0$ is the cavity (linear) refractive index
\item $n_2$ is the cavity intensity-dependent (nonlinear) refractive index
\item $A_{\textrm{eff}}$ is the effective mode area
\item $\Delta f_{\textrm{FSR}}$ is the free spectral range of the resonator
\item $P_{\textrm{in},i}$ is the input laser power in direction $i$
\item $P_{\textrm{circ},i}$ is the power in the circulating cavity mode in direction $i$
\end{enumerate}
The normalised detuning for propagating direction $i$ is the actual laser detuning ($\delta_i = \omega_{\textrm{las},i} - \omega_0$) divided by the loaded cavity linewidth ($\gamma=\gamma_0+\kappa$):
\begin{equation}
\Delta_i = \frac{-\delta_i}{\gamma}
\end{equation}
The characteristic power at which the Kerr effect is observed is given by:
\begin{equation}
P_0 = \frac{\pi n_0 A_{\textrm{eff}}}{n_2 Q \mathcal{F}_0}
\end{equation}
where the intrinsic cavity finesse, $\mathcal{F}_0=\pi \Delta f_{\textrm{FSR}}/\gamma_0$ and the loaded cavity \textit{Q}-factor, $Q=\omega_0/2 \gamma$.
\par
The normalised input power in direction $i$ is then given by:
\begin{equation}
\tilde{p}_i = \frac{\eta P_{\textrm{in},i}}{P_0}
\end{equation}
where the coupling efficiency, $\eta = 4\kappa \gamma_0 / \gamma^2$.
\par
Finally, the circulating power in the cavity mode in direction $i$ is given by:
\begin{equation}
p_i = \frac{2 \pi P_{\textrm{circ},i}}{\mathcal{F}_0 P_0}
\end{equation} 

\bibliographystyle{IEEEtran}
\bibliography{IEEEabrv,logic}

\end{document}